# Nonlinear statistical coupling

Kenric P. Nelson[a*] and Sabir Umarov[b]

[a] *Raytheon Company, Woburn, MA 01801 USA*
[b] *Dept. of Mathematics, Tufts University, Medford, MA, 02155 USA*

*Abstract* — By considering a nonlinear combination of the probabilities of a system, a physical interpretation of Tsallis statistics as representing the nonlinear coupling or decoupling of statistical states is proposed. The escort probability is interpreted as the *coupled probability*, with $Q = 1 - q$ defined as the degree of nonlinear coupling between the statistical states. Positive values of $Q$ have coupled statistical states, a larger entropy metric, and a maximum *coupled-entropy* distribution of compact-support *coupled-Gaussians*. Negative values of $Q$ have decoupled statistical states and for $-2 < Q < 0$ a maximum coupled-entropy distribution of heavy-tail coupled-Gaussians. The conjugate transformation between the heavy-tail and compact-support domains is shown to be $\hat{Q} = \frac{-2Q}{2+Q}$ for coupled-Gaussian distributions. This conjugate relationship has been used to extend the generalized Fourier transform to the compact-support domain and to define a scale-invariant correlation structure with heavy-tail limit distribution. In the present paper, we show that the conjugate is a mapping between the source of nonlinearity in non-stationary stochastic processes and the nonlinear coupling which defines the coupled-Gaussian limit distribution. The effects of additive and multiplicative noise are shown to be separable into the *coupled-variance* and the coupling parameter $Q$, providing further evidence of the importance of the generalized moments.

## 1. Introduction

Part of the foundation of probability and statistics is the consideration of mutually exclusive states with probabilities which sum to one, representing all possible configurations of the system. Nevertheless, there are instances where the states of a physical system are complicated by either incomplete statistics [1], fluctuations in the statistical parameters defining the system [2-4], long-range correlations [5-7] or the nonlinear dynamics of the system [8]. In each of these cases models have been defined which lead to an escort probability, which modifies the probabilities by a power-law, a non-additive entropy function, and a generalization of the Gaussian distribution, known as the *q*-Gaussian [9, 10]. The methods of *q*-statistics have been shown to provide techniques for analysis of turbulence [11], communication signals [12, 13], dynamics of the solar wind [14], neural networks [15] and other complex phenomena.

In this paper we attempt to provide a definition for the parameter *q* which quantifies a physical property, which has been cited by some critiques as a deficiency within the *q*-statistics framework [16-19]. An interpretation of *q*-statistics as a model of nonlinear coupling between the statistical states of a system is introduced in Section 2. The mathematical symmetry of this approach makes evident a conjugate relationship between heavy-tail and compact-support *q*-Gaussians which is defined is section 3. The conjugate relationship is shown in section 4 to relate the source of nonlinearity in a stochastic process to the degree of nonlinear coupling in the stationary distribution. Section 5 provides concluding remarks and suggestions for future inquiry.

## 2. An interpretation of *q*-statistics using nonlinear coupling

To begin, we propose a model of the nonlinear coupling between statistical states, which results in a definition of the degree of *nonlinear statistical coupling*. Suppose that the dynamics of a system are such that the individual probabilities of a system $P_i$ cannot be treated in isolation. Instead, consider this individual probability with a nonlinear combination of all the probabilities of the system, $P_i \left( P_1^Q P_2^Q \cdots P_{j \neq i}^Q \cdots P_N^Q \right)$, where N is the number of states and $Q$ is the nonlinear coupling strength between the probabilistic states. Normalization of this nonlinear combination leads to the following definition

[*] Corresponding author, Phone 603-508-9827;
E-mail address: kenric.nelson@ieee.org (K. Nelson)

**Definition 1.** *The coupled probability*

$$P_{Q,i} \equiv \frac{P_i \prod_{\substack{j=1 \\ j \neq i}}^{N} P_j^Q}{\sum_{i=1}^{N} P_i \prod_{\substack{j=1 \\ j \neq i}}^{N} P_j^Q} = \frac{P_i^{1-Q}}{\sum_{i=1}^{N} P_i^{1-Q}}. \tag{1}$$

The effect of coupling with a zero probability state is eliminated by the normalization. This probability is equivalent to the escort probability of $q$-statistics with the parameter translated by $Q = 1-q$. While only one system-wide coupling parameter is considered here, the model is suggestive of a more general situation in which the coupling strength between states is different. Following from the coupled-probability, the functions of $q$-statistics are expressed using the translated parameter, and the phrase *coupled-{function}* will be used given the definition of $Q$ as a coupling coefficient. We will show that the improved mathematical symmetry and direct modeling of a physical property provide advantages for the conceptual representation of stochastic systems impacted by sources of nonlinearity. Details of the translated $q$-calculus are described in [20] and has been utilized directly by Topsoe in [21], Martinez in [22], and Wang in [23] and as complementary analysis in for example [4, 5]. The translation between $Q$ and $1-Q$ is equivalent to the multiplicative dual of $q$-statistics, which is $Q_m = \frac{Q}{1-Q}$ in the translated form. For continuous variables, the *coupled probability density* is $f_Q(x) \equiv \frac{f^{1-Q}(x)}{\int_{-\infty}^{\infty} f^{1-Q}(x)dx}$, where $f(x)$ is the probability density. The moments of a coupled system are defined using the coupled probability [24]. The required deformation from the standard moments depends on the asymptote of the function, which given $\lim_{|x| \to \infty} f(x) = |x|^{\alpha/\gamma}$, is a function of $\alpha$, the power within the exponential family ($\alpha = 1$ is exponential and $\alpha = 2$ is Gaussian) and of $Q$, the degree of deformation from the exponential family.

**Definition 2.** *The $n^{th}$ coupled-moment for a function with asymptote* $\lim_{|x| \to \infty} f(x) = |x|^{\alpha/\gamma}$ *is*

$$\left\langle X^n \right\rangle_{nQ/\alpha} \equiv \int_{-\infty}^{\infty} x^n f_{nQ/\alpha}(x)dx. \tag{2}$$

While it may seem inappropriate for the definition to depend on properties of the distribution, we will show that this facilitates the separation of non-stationary influences generating power-law characteristics from the stationary influences characteristic of Gaussian distributions. This definition suggests a coupled-mean of $\mu_{Q/2}$ for the Gaussian, but we will use the more common definition $\mu_Q$, since the discussion is limited to symmetric distributions. The coupled-variance for the Gaussian is $\sigma_Q^2$.

Claude Shannon [25] demonstrated that the logarithmic function was the proper analytical tool for measuring the information of a probability distribution. Implicit in this metric is the centrality of the exponential family within probability theory. Constantino Tsallis and other investigators [8, 26-28] have sought to extend entropic analysis to systems with power-law characteristics. The deformed logarithmic function is defined as $\ln_Q(x) \equiv \frac{x^Q - 1}{Q}$ and provides a smooth transition from logarithmic ($Q \to 0$) to linear ($Q = 1$), quadratic ($Q = 2$), ... power-law functions. Its inverse is the generalized exponential $e_Q^x \equiv (1+Qx)_+^{1/Q}$, where $(y)_+ \equiv \max(0, y)$. The generalized surprisal [29] or information gain from the occurrence of an event with probability $p_i$ is defined as

$$coupled - surprisal \equiv \ln_Q \left( 1/p_i \right) = -\ln_{-Q}(p_i) = \frac{p_i^{-Q} - 1}{Q}. \tag{3}$$

For positive (negative) values of $Q$ the surprisal or information metric increases (decreases) in size. At $Q = -1$ the coupled-surprisal is a linear function of $p_i$. For $q > -1$ the function is concave and for $Q < -1$ the function is convex. The equality $\ln_Q x^{-1} = -\ln_{-Q} x$ is an example of the additive dual $Q_a = -Q$. The coupled-entropy is the average coupled-surprisal $S_Q = \left\langle \ln_Q \left( 1/p_i \right) \right\rangle = \frac{1}{Q} \left( -1 + \sum_{i=1}^{N} p_i^{1-Q} \right)$ and is non-additive for independent systems, $S_Q(A,B) = S_Q(A) + S_Q(B) +$ $QS_Q(A)S_Q(B)$. The entropy is extensive for $Q_{ent}$ if $\lim_{N\to\infty} S_{Q_{ent}}(N) \sim cN$. Examples of systems with severe correlations [7, 30] have been shown to be extensive for $0 < Q_{ent} < 1$. The positive value of $Q$ increases the entropy measure, compensating for the loss of information due to the correlations. While examples of $Q_{ent} < 0$ are not as common, entangled quantum systems, which can have super-additive entropy [31, 32], may be an application; and this domain is useful in modeling the influence of risk, beliefs, or other forms of a-priori information gain [33-35].

Using the first and second coupled-moments, $\mu_Q$ and $\sigma_Q^2$ for $\alpha = 2$, as constraints leads to the coupled-Gaussian as the maximum coupled-entropy distribution

$$G_Q(x; \mu_Q, \sigma_Q^2) = \frac{1}{\sqrt{2+q}C_Q \sigma_Q} \exp_Q\left[-\frac{(x-\mu_Q)^2}{(2+Q)\sigma_Q^2}\right] = \frac{1}{\sqrt{2+Q}C_Q \sigma_Q}\left[1 + \left(\frac{2Q}{2+Q}\right)\left(\frac{-(x-\mu_Q)^2}{2\sigma_Q^2}\right)\right]_+^{\frac{1}{Q}}, \quad (4)$$

where $C_Q$ is the normalization, and the term $\frac{2Q}{2+Q}$ will be shown in the next section to be an important transformation. The coupled-Gaussian distribution is thus characteristic of systems in which the probabilities of the discrete states of the system are coupled with a strength of $Q$. In the case of zero coupling strength $(Q \to 0)$ the distribution is the traditional Gaussian, common throughout statistical analysis. For positive values of $Q$, the nonlinear coupling between the states and the increased cost of rare events defined by the $Q$-surprisal strengthens the rate of decay, resulting in a distribution with compact support. In this case, the probability is zero for $|x - \mu_Q| \geq \sqrt{\frac{(2+Q)}{Q}}\sigma_Q$. For $Q = 1$ the distribution is parabolic and as $Q \to \infty$ the distribution approaches uniformity. The compact-support distributions are representative of the gains in information due to reductions in fluctuations which eliminate the probability of rare states. These distributions can also arise in finite domain systems in which correlations or losses of information increase the probability of states near the boundary [5, 6, 36]. For negative values of $Q$, the decoupled states and decreased coupled-surprisal weakens the rate of decay, resulting in a heavy-tail distribution. Between $-2 < Q \leq -\frac{2}{3}$ the classic variance is divergent, but the coupled-variance is finite. $Q = -1$ is the Cauchy distribution. Beyond $Q < -2$ the distribution cannot be normalized; i.e. $C_Q$ is divergent, and will be shown to be a natural boundary of physical systems. The heavy-tail distributions are often representative of losses in information due to correlations or non-stationary fluctuations.

## 3. Conjugate Pairs of Heavy-tail and Compact-Support distributions

A sequence of $Q$ parameters has been defined in the context of repeated application of the generalized Fourier transform [37]. The $n^{th}$ sequence value of $Q$ is

$$z_n(Q) \equiv Q_n \equiv \frac{2Q}{2+nQ} = \left[\frac{1}{Q} + \frac{n}{2}\right]^{-1} \quad n = 0, \pm 1, \pm 2, \ldots \quad (5)$$

where positive values of $n$ represent application of the $Q$-FT and negative values represent application of the inverse $Q$-FT. This sequence is also related to the $n^{th}$ integral (derivative) of the coupled-Gaussian, which does not preserve the coupled-Gaussian structure but does have an asymptotic power of $\frac{2}{Q} + n$.

The $Q$-sequence can be used to define a mapping between the heavy-tail $Q$-Gaussians between $-2 < Q < 0$ and the compact-support coupled-Gaussians between $0 < Q < \infty$. For guidance in this mapping, first consider the Fourier transform for a power-law function, which the heavy-tail coupled-Gaussians approach asymptotically as x goes to infinity. For $-2 < Q < 0$, the tail of the coupled-Gaussian approaches a power-law

$$e_Q^{-\beta x^2} \sim (-Q\beta x^2)^{\frac{1}{Q}} \sim O(x^{\frac{2}{Q}}), \quad x \to \infty \quad (6)$$

The Fourier transform of a power law is also a power law with the following form [38, 39]

$$|x|^{\frac{2}{Q}} \Leftrightarrow \sqrt{\frac{2}{\pi}}\Gamma(\frac{2}{Q}+1)\sin(-\frac{\pi}{Q})|\omega|^{-(\frac{2}{Q}+1)} \quad \text{for } -2 < Q < 0 \quad (7)$$

where $\omega$ is the dual variable (or frequency). In the frequency domain the power is $-(\frac{2}{Q}+1)$ which rearranged in the form of a coupled-Gaussian, shows the following relationship with the $Q$-sequence

$$\left|\omega^2\right|^{-\frac{2+Q}{2Q}} = \left|\omega^2\right|^{-\frac{1}{Q_1}} \tag{8}$$

So the Fourier transform of a power law is suggestive of an important mapping between a $Q$-Gaussian and a $-Q_1$-Gaussian. In fact the mapping between $G_Q \leftrightarrow G_{-Q_1}$ constitutes a conjugate dual relating the heavy-tail $Q$-Gaussians to the compact-support $Q$-Gaussians. The numeral 2 generalizes to $0 < \alpha \leq 2$ for the $(Q, \alpha)$- distributions [6, 8-10], $G_{Q,\alpha}(x) \equiv a e_Q^{-\beta|x|^\alpha} = a\left(1 - Q\beta|x|^\alpha\right)^{1/Q}$ which is the coupled-Fourier image of the generalization of alpha-stable Lévy distributions. More generally, for the $(Q, \alpha)$- distribution, the compact-support range $Q > 0$ is mapped to the heavy-tail range $-\alpha < -Q_1 < 0$ and vice-versa. This relationship leads to the definition for the *conjugate dual*.

**Definition 3.** *The conjugate dual is defined as*

$$\hat{Q}^{(\alpha)} \equiv -z_{(\alpha,1)}(Q) = \frac{-\alpha Q}{\alpha + Q}. \tag{9}$$

The inverse of the conjugate is the same function: $\hat{z}_\alpha^{-1}(\hat{Q}) = \frac{-\alpha \hat{Q}}{\alpha+\hat{Q}} = \frac{-\alpha(-Q_1)}{\alpha+(-Q_1)} = Q$. Since $\alpha$ is generally known by the context of a problem, this general conjugate will be shortened to $\hat{Q}$. If $\alpha = 1$ the conjugate is a conjunction of the additive and multiplicative duals $\hat{Q} = q_m(q_a)$. A set of conjugate $q$-Gaussian pairs with $\sigma_Q^2 = \sigma_{\hat{Q}}^2 = 1$ is shown in Figure 1. On the left of Figure 1 are the compact-support distributions which proceed from the Gaussian distribution for values close to zero and converge to the uniform distribution as $Q$ goes to infinity. On the right are the conjugate heavy-tail distributions, which converge to an infinitesimal density at $Q = -2$. If $\beta$ is invariant the compact-support distribution converges to a delta function as $Q$ goes to infinity. This conjugate relationship has been utilized to extend the $Q$-Fourier transform to the compact-support domain [20] and to extend scale-invariant correlation models to the heavy-tail domain [6].

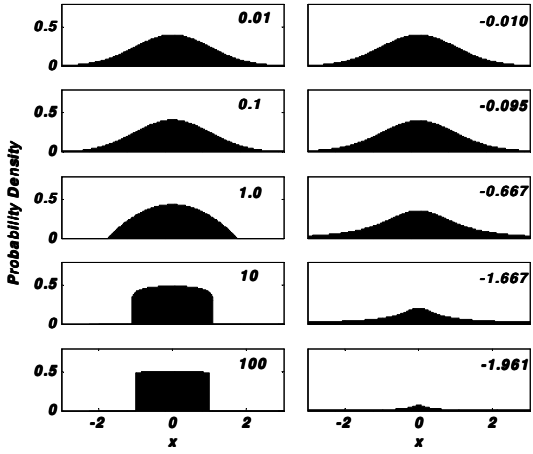

Figure 1: The conjugate pairs of $Q$-Gaussian distributions with $\sigma_Q^2 = 1$. The value of $Q$ is inset in each figure. On the left are the compact-support $Q$-Gaussians with $Q > 0$. On the right are the conjugate heavy-tail distributions with $-2 < Q < 0$. The conjugate distribution has $\hat{Q} = -Q_1 = \frac{-2Q}{2+Q}$. As $Q$ approaches infinity the distribution tends toward the uniform while its conjugate at -2 is spread infinitesimally to the tails of the distribution.

In the next section, we show that the nonlinear source generating a $Q$-Gaussian limit is equal to $\hat{Q} = -Q_1$ and that the coupled-moment terms of the $Q$-Gaussian $\frac{-(x-\mu_Q)^2}{2\sigma_Q^2}$ are independent of the nonlinear source. As noted in (4) the remaining term $\frac{2Q}{2+Q}$ is $Q_1 = -\hat{Q}$. A revised representation for the coupled-Gaussian is used so that the coupled-exponent is not dependent on the source of nonlinearity

**Definition 4.** *Given the coupled-mean $\mu_Q$ and coupled-variance $\sigma_Q^2$, the $(Q, Q_1)$-Gaussian is defined as*

$$\exp_{(Q,Q_1)}\left(\frac{-(x-\mu_Q)^2}{2\sigma_Q^2}\right) \equiv \left(1 + Q_1\left(\frac{-(x-\mu_Q)^2}{2\sigma_Q^2}\right)\right)^{1/Q} = \left(1 + \left(\frac{2Q}{2+Q}\right)\left(\frac{-(x-\mu_Q)^2}{2\sigma_Q^2}\right)\right)^{1/Q}. \tag{10}$$

A more general definition with $Q_{(\alpha,1)} = \frac{\alpha Q}{\alpha+Q}$ is expected to be applicable for $0 < \alpha \leq 2$.

## 4. Sources of nonlinear statistical coupling

The conjugate transformation provides a key insight into the relationship between the nonlinear statistical coupling of the metastable distribution, known as the $Q$-stationary distribution ($Q_{stat}$) and the nonlinear source generating the non-stationary process, denoted $Q_{source}$. Table 1 provides a synopsis of applications using this interpretation. In the following theorem we provide a proof of this relationship for a stochastic process with multiplicative noise [40]. Further, the coupled-variance is only dependent on the additive noise.

**Theorem 1.** *Let a stochastic process $X_t$ with multiplicative noise, be defined by the stochastic differential equation*

$$dX_t = f(X_t)dt + g(X_t)(2MdW_t^{(m)}) + (2AdW_t^{(a)}) \tag{11}$$

*where $dW_t^{(m)}$ and $dW_t^{(a)}$ are independent Wiener processes which define the multiplicative (m) and additive (a) noise, and M and A are the amplitudes of each noise source. And let $f(x) = -\tau g(x)g'(x) = -V'(x)$, where $V(x)$ is a potential function specifying the influence of the multiplicative noise on the diffusion $D(x)$. Then the probability density $p_X(x,t)$ for this system has a limit solution of $p_X(x) = \lim_{t\to\infty} p_X(x,t) \propto \exp_{(Q,Q_1)}\left[-\frac{g(x)^2}{2\sigma_Q^2}\right]$, with $\sigma_Q^2 = \frac{A}{\tau}$ and $Q = \frac{-2M}{\tau + M} = \hat{Q}_{source}$, where $Q_{source} = D'(x)/V'(x)$.*

*Remarks.* This limit solution was discussed by Anteneodo in [40]; here we show that $Q_{stat} = \hat{Q}_{source}$ and that the coupled-variance only depends on the additive noise. If $g(x) \approx x - \mu_q + O(x^2)$ then $p_X(x)$ is approximated by a coupled-Gaussian. When the multiplicative noise $M$ is zero, we have $Q = 0$, and the approximate stationary solution with additive noise is a Gaussian distribution in accordance with the classical theory. By using Definition 4 for the coupled-Gaussian the coupled-exponent is independent of the multiplicative noise. Importantly, the deterministic, additive noise, and nonlinearity of a stochastic process are separable into the coupled-mean, coupled-variance, and $Q$ parameter of the coupled-Gaussian distribution. This is further evidence that the coupled-moments quantify important physical properties of complex systems with power-law characteristics.

*Proof.* The probability density $p_X(x,t)$ is the solution to the Fokker-Plank equation [40]

$$\frac{\partial p_X(x,t)}{\partial t} = -\frac{\partial}{\partial x}[J(x)p_X(x,t)] + \frac{\partial^2}{\partial x^2}[D(x)p_X(x,t)] \tag{12}$$

where $J(x)$ is the drift and $D(x)$ is the diffusion:

$$J(x) \equiv f(x) + Mg(x)g'(x)$$
$$D(x) \equiv A + M[g(x)]^2 \tag{13}$$

Since $g(x)$ is related to the potential function by $V'(x) = \tau g(x)g'(x) \to V(x) = \frac{\tau}{2}[g(x)]^2$, the diffusion rate and the potential function are related by the expression

$$D(x) = A + \frac{2M}{\tau}V(x). \tag{14}$$

The magnitude of the nonlinear source is $Q_{source} = D'(x)/V'(x) = \frac{2M}{\tau}$. The stationary solution $p_X(x)$ for (12) is a member of the coupled-exponential family [41, 42], and as shown in [40]

$$p_X(x) = \lim_{t\to\infty} p_X(x,t) \propto \exp_{(Q,Q_1)}\left[-\frac{g(x)^2}{2\sigma_Q^2}\right]. \tag{15}$$

By using the coupled-variance $\sigma_Q^2 = \frac{A}{\tau}$, and the coupling parameter $Q_{stat} = \frac{-2M}{\tau + M} = \hat{Q}_{source}$, the separation between the effects of the additive and multiplicative noise is evident, completing the proof. □

While the theorem has specific conditions, the applicability of the mapping between a nonlinear source and the stationary $Q$-Gaussian appears to be broader. Another well-established source of the coupled-Gaussian distribution is the ratio of a normal random variable and a chi-square variable, because of its equivalence to the student-T distribution. Given the degree of freedom $v$ the Student-T [43] distribution is $f(x) = \frac{\Gamma(\frac{v+1}{2})}{\sqrt{v\pi}\Gamma(\frac{v}{2})}\left(1+\frac{x^2}{v}\right)^{-\left(\frac{v+1}{2}\right)}$, which can be equated to a heavy-tail $(Q, Q_1)$-Gaussian $A_Q e_{(Q,Q_1)}^{-x^2/2}$ with $Q = -\left(\frac{2}{v+1}\right)$, $Q_{source} = \hat{Q} = \frac{2}{v}$ and $\sigma_Q^2 = 1$. A related result is a generalized central-limit for coupled-Gaussians by Vignat and Plastino [44] in which independent variables by are weighted by a chi-square distribution. Superstatistic models [2, 4, 45] based on fluctuations of the inverse-width coefficient $\beta = [(2+Q)\sigma_Q^2]^{-1}$ have been shown to be appropriately modeled by coupled-Gaussians either exactly for chi-square distributions of $\beta$ or approximately for small fluctuations with more general distributions. For Type-B superstatistics the normalization of the Boltzmann factor $Z(\beta)$ is included within the fluctuation average, $p(E) = \int_0^\infty f(\beta) Z^{-1}(\beta) e^{-\beta E} d\beta$, where $E$ is energy. The relative variance $\frac{\langle \beta \rangle^2 - \langle \beta^2 \rangle}{\langle \beta \rangle^2}$ is equal to $Q_{source}$ and $Q_{stat} = \hat{Q}_{source}$. While Type-A superstatistics, which normalizes the distribution after averaging over $\beta$ can be equated to Type-B, the relative variance is then equal to $-Q_{stat}$ rather than $Q_{source}$, which is a less favorable model.

*Table 1:* Sources of nonlinear statistical coupling with stationary distributions which are heavy-tail Q-Gaussians.

| Physical Source | Nonlinear Source $0 \leq Q_{source} < \infty$ | Nonlinear Coupling $-2 \leq Q_{stationary} = \hat{Q}_{source} < 0$ | Physical Parameters |
|---|---|---|---|
| Degree of Freedom for Student-T | $2/v$ | $\frac{-2}{v+1}$ | $v$ - degree of freedom |
| Lévy-like anomalous diffusion [46] | $2/\gamma_L$ | $\frac{-2}{\gamma_L + 1}$ | $\gamma_L$ - index of Lévy distribution |
| Multiplicative Noise [40] | $2M/\tau$ | $\frac{-2M}{\tau + M}$ | $M$ – multiplicative noise $\tau$ - amplitude of potential |
| Type-B Super-statics [2, 45] | $relVar = \frac{\langle \beta^2 \rangle - \langle \beta \rangle^2}{\langle \beta \rangle^2}$ | $\frac{-2 relVar}{2 + relVar}$ | $\beta$ – fluctuating parameter |

The connection between a nonlinear source and long-range correlation is less direct since the transition to the coupled-Gaussian as a limit distribution is established via an asymptotic or strictly scale-invariant pattern of linear correlations between the system states. Nevertheless, the results of Hanel, *et. al.* [6], where the probabilities for a N size system are $p_n^N = \binom{N}{n} r_n^N$ and the scale-invariant correlation is established by applying the Leibtniz rule $r_n^{N-1} = r_n^N + r_{n+1}^N$, can be interpreted as requiring a nonlinear source $Q_{source}$ which for $N \to \infty$ approaches the $(Q_{\lim}, -Q_{source})$-Gaussian distribution. One such model for heavy-tail distributions is

$$r_n^N = \frac{B\left(\frac{1}{Q_{source}} + n, \frac{1}{Q_{source}} + N - n\right)}{B(\frac{1}{Q_{source}}, \frac{1}{Q_{source}})}; \quad -2 < Q_{\lim} = \hat{Q}_{source} < 0 \qquad (16)$$

Where $B(\alpha, \beta) = \int_0^1 x^\alpha (1-x)^\beta dx$ is the beta function and a mapping function is required between the domain of the beta function and the domain of the limit distribution. This solution draws directly upon the *q*-conjugate originally defined in [20]. The scale-invariant correlation has also been used to model compact-support processes; however, the results are based on requiring a finite-domain solution and applying an appropriate correlation structure for coupled-Gaussian limit distribution. A purely conjugate relationship with (16) would require $Q_{\lim}$ and $Q_{source}$ to be swapped, in which case the system would have a source of reduced variation resulting in the compact-support limit distribution. Further investigation of the differences between these types of compact-support limit distributions would be of interest.

## 5. Conclusion

In this investigation of non-stationary stochastic processes, we have sought the simplest mathematical expressions for $q$-statistics. The result is a symmetrical definition for $q$-statistics in which $Q = 1 - q \to 0$ represents non-existence of the generalized form. From this point of view, $Q$ can quantify a physical property, which we propose is nonlinear statistical coupling based on a definition for the coupled probability. A conjugate transformation between the compact-support domain ($0 < Q < \infty$) and the heavy-tail domain ($-\alpha < Q < 0$) is defined by $\hat{Q} = \frac{-\alpha Q}{\alpha + Q} = -\left[ \frac{1}{Q} + \frac{1}{\alpha} \right]^{-1}$ for the $(Q, \alpha)$-distribution, which is a coupled-Gaussian for $\alpha = 2$. The conjugate has several important properties. The power terms of a heavy-tail distribution $Q$-Gaussian ($-\frac{2}{Q}$) and its Fourier image ($-\frac{2}{\hat{Q}}$) are related by the conjugate. The relationship is its own inverse: $\hat{z}(\hat{Q}) = Q$. The conjugate has been used to define a coupled-Fourier transform in the compact-support domain and to define a conjugate coupled-Fourier transform between the heavy-tail and compact-support domains [20]. We demonstrate that the nonlinearity due to multiplicative noise in a stochastic process is related to the nonlinear statistical coupling of the stationary distribution via the conjugate, $Q_{stat} = \hat{Q}_{source}$ and that this relationship is consistent for other sources of heavy-tail coupled-Gaussians. The importance of the coupled-variance is strengthened by showing that it measures the additive noise properties of a stochastic process, independent of the multiplicative noise.

This paper has focused on the conjugate relationship between a nonlinear source and the $Q$-stationary distribution for the one-dimensional heavy-tail domain. The compact-support domain requires a separate discussion to describe the differences between systems in which a nonlinear source with negative coupling strength induces a restricted domain with positive nonlinear statistic coupling and those with a restricted domain in which correlation or fluctuations increase the probability of states near the finite boundary. The former case is expected to be the reciprocal of the heavy-tail examples discussed here, while the latter case will have a different transformation between the source and stationary distribution. Of additional interest are applications for $\alpha \neq 2$ and/or dimensions other than one. In particular, the $Q$-sensitivity and $Q$-relaxation parameters, the counterparts to $Q$-stationary parameter, forming a triplet of physical parameters for complex systems, are related to coupled-exponentials ($\alpha = 1$) [10, 14], and would be interesting to revisit within the context of mapping the nonlinear source and distribution parameters.

### Acknowledgment

The authors are grateful for conversations with Fleming Topsoe and Constantino Tsallis. P.K. Rastogi provided valuable comments during preparation of the manuscript.